\documentclass{article}

\usepackage{amsmath}
\usepackage{amssymb}
\usepackage{listings}
\usepackage{graphicx}
\usepackage{hyperref}

\title{An Extensible Julia Toolkit for Symmetry-Aware Dual Space Phasing in Arbitrary Dimensions}

\author{Pavel Kalugin\footnote{Laboratoire de Physique des Solides, CNRS, Universit\'e Paris-Sud, Universit\'e Paris-Saclay, F-91405 Orsay, France. E-mail: \texttt{kalugin@lps.u-psud.fr}}}

\begin{document} 
\maketitle 

\begin{abstract}
We present an open-source Julia-based software toolkit for solving the phase problem using dual-space iterative algorithms. The toolkit is specifically designed for aperiodic crystals and quasicrystals, supporting general space group symmetries in arbitrary dimensions. A key feature is the symmetry-breaking anti-aliasing sampling scheme, optimized for computational efficiency when working with strongly anisotropic diffraction data, common for quasicrystals. This scheme avoids sampling redundancy caused by symmetry constraints, imposed during phasing iterations. The toolkit includes a reference implementation of the charge flipping algorithm and also allows users to implement custom phasing algorithms with fine-grained control over the iterative process. 
\end{abstract}

\section{Introduction}
The development of dual-space iterative algorithms was preceded by several milestones. The “shake-and-bake” method \cite{miller1994snb} first demonstrated an iterative approach alternating modifications in real and reciprocal space. Then, with the principle of minimal charge'' \cite{elser1999x}, direct manipulation of atomic positions was replaced by refinement of the continuous charge-density distribution. This was critical for the treatment of quasicrystals, where the density in the high-dimensional unit cell is concentrated on “atomic surfaces” rather than at discrete atomic sites. The final piece of puzzle was the replacement of explicit maximization of the global minimum of the charge density by the flow of a discrete dynamical system --- the charge-flipping procedure \cite{oszlanyi2004ab}. Since then, dual-space iterative methods have been widely applied to the phase problem in X-ray crystallography (see \cite{palatinus2013charge} for review). Despite their practical success, the convergence properties of these methods remain poorly understood (although systematic studies of their convergence properties are beginning to emerge, e.g., \cite{elser2018benchmark}). Consequently, existing dual-space phasing programs, such as SUPERFLIP \cite{palatinus2007superflip}, rely heavily on heuristic strategies. Furthermore, the role of symmetry constraints during iteration is still a matter of debate. These issues highlight the need for an open platform facilitating experimentation with dual-space phasing algorithms. Such a platform should manage routine tasks like charge density sampling and symmetry handling, while allowing flexible implementation of iterative phasing schemes.
\par
We present a software toolkit, implemented in Julia \cite{bezanson2017julia}, designed for rapid prototyping and testing of novel charge-flipping algorithms. Julia was selected to enable a modular architecture without the constraints of the traditional two-language paradigm, in which computational kernels are implemented in a low-level language (e.g. C or Fortran) and accessed via a high-level interface (e.g. MATLAB or Python). Clearly, such interfaces do not allow for modification of the core algorithms. Julia addresses this {\em two-language problem} by combining high-level expressiveness with performance comparable to compiled languages. Moreover, the modern type system of Julia is well suited to represent complex mathematical objects, while its multiple dispatch model supports a clean and extensible software design. The toolkit comprises the main package \lstinline{ChargeFlipPhaser} \cite{ChargeFlipPhaser} and the auxiliary package \lstinline{SpaceGroups} \cite{SpaceGroupsJL}.
\section{Data preparation}
The toolkit does not impose a fixed data input format. Instead, the user provides a Julia script that performs the following tasks:
\begin{itemize}
\item \textbf{Symmetry and lattice specification:} The user defines the symmetry group either by referencing a pre-defined group from the toolkit's library or by explicitly listing its generators. The parameters of the reciprocal lattice are specified by giving the basis of the $\mathbb{Z}\text{-module}$ of Bragg peaks. For example, for an icosahedral quasicrystal the user should provide a real-valued $3 \times 6$ matrix.
\item \textbf{Bragg peak data entry:} Diffraction peaks are added individually, by specifying their indices and intensities. Peaks related by a symmetry operation to previously added reflections are ignored, and attempts to add peaks that should be extinct under the specified symmetry result in an exception.
\item \textbf{Form factor specification:} Form factors used to scale the diffraction amplitudes must also be supplied (see below).
\end{itemize} 
\subsection{Form factors}\label{sec:formfactors}
\begin{figure}[ht] %
    \begin{center}
        \includegraphics[width=\textwidth]{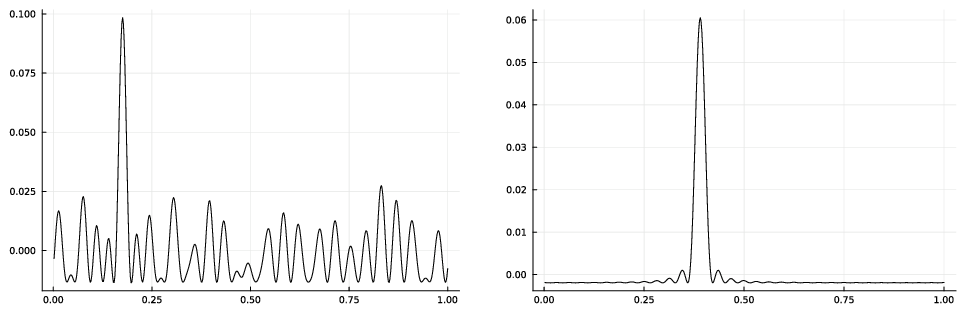} 
    \end{center}
    \caption{Effect of applying a windowing function to normalized diffraction amplitudes. The charge density profiles were obtained using the reference implementation of the algorithm on simulated data for a one-dimensional crystal with a single atom per unit cell and 32 Bragg reflections (vertical scale arbitrary). Left: Reconstruction using raw normalized amplitudes. The algorithm converges to the optimal solution under its objective function --- maximizing the minimal density --- as evidenced by multiple local minima all reaching the same level. However, this solution does not correspond to the correct one (a sinc-shaped peak). Right: Result obtained with a Bartlett window applied to the amplitudes. In this case, the algorithm converges to the correct solution.} 
    \label{fig:formfactors}
\end{figure}

Form factors are critical for improving the convergence behaviour of charge flipping algorithms. They are specified as arrays of scaling coefficients applied to the diffraction amplitudes. These coefficients must vary smoothly in reciprocal space to preserve the integrity of reconstructed atomic positions. Typically, the applied form factor is a product of two components:
\begin{itemize}
    \item \textbf{A sharpening factor}, used to normalize the structure factors $F$ into $E$-values, following the standard approach in charge flipping \cite{oszlanyi2008charge}. Normalization may be achieved by the Wilson plot technique \cite{shmueli2001international}, or computed directly using tabulated atomic form factors (included in the toolkit, following \cite{waasmaier1995new}) and Debye-Waller factors estimated from analogous structures.
   \item \textbf{A regularization factor}, which mitigates density oscillations caused by the finite cutoff in reciprocal space. These oscillations may otherwise lead the algorithm to converge to artefactual solutions, since charge flipping seeks to maximize the global minimum of the charge density. This risk is illustrated in Figure~\ref{fig:formfactors}, left panel. Regularization via a smooth windowing function suppresses such artefacts, as seen in the right panel of the same figure. The toolkit provides a built-in windowing function for this purpose, based on the autocorrelation of a spherical ball indicator function in reciprocal space.
\end{itemize}
\section{Charge sampling strategy}
The freedom to choose the grid for charge density sampling is often left unexploited. Most existing dual-space phasing software employs only Cartesian grids expressed in a fixed basis:
\begin{equation}
    \mathbf{x}_{m_1,\dots,m_d} =\sum_{i=1}^d \frac{m_i}{n_i} \mathbf{e}_i, 
    \label{eq:cartesian_grid}
\end{equation}
where $\mathbf{e}_i$ are the unit cell basis vectors of a $d\mbox{-dimensional}$ lattice, $n_i$ are positive integers, and $1 \le m_i \le n_i$. This choice is suboptimal for at least two reasons: symmetry considerations and aliasing avoidance.
\subsection{Symmetry breaking}
\begin{figure}
    \centering
    \begin{minipage}[b]{0.48\linewidth}
        \centering
        \includegraphics[width=\linewidth]{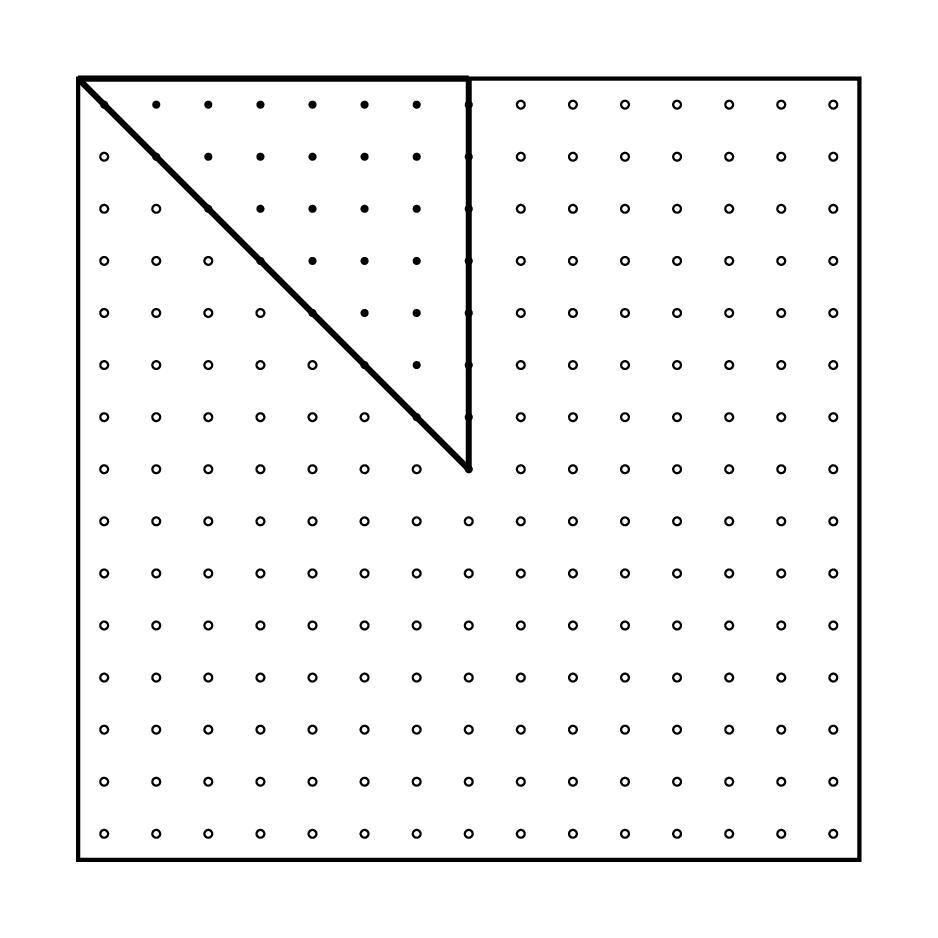}
    \end{minipage}
    \hfill
    \begin{minipage}[b]{0.48\linewidth}
        \centering
        \includegraphics[width=\linewidth]{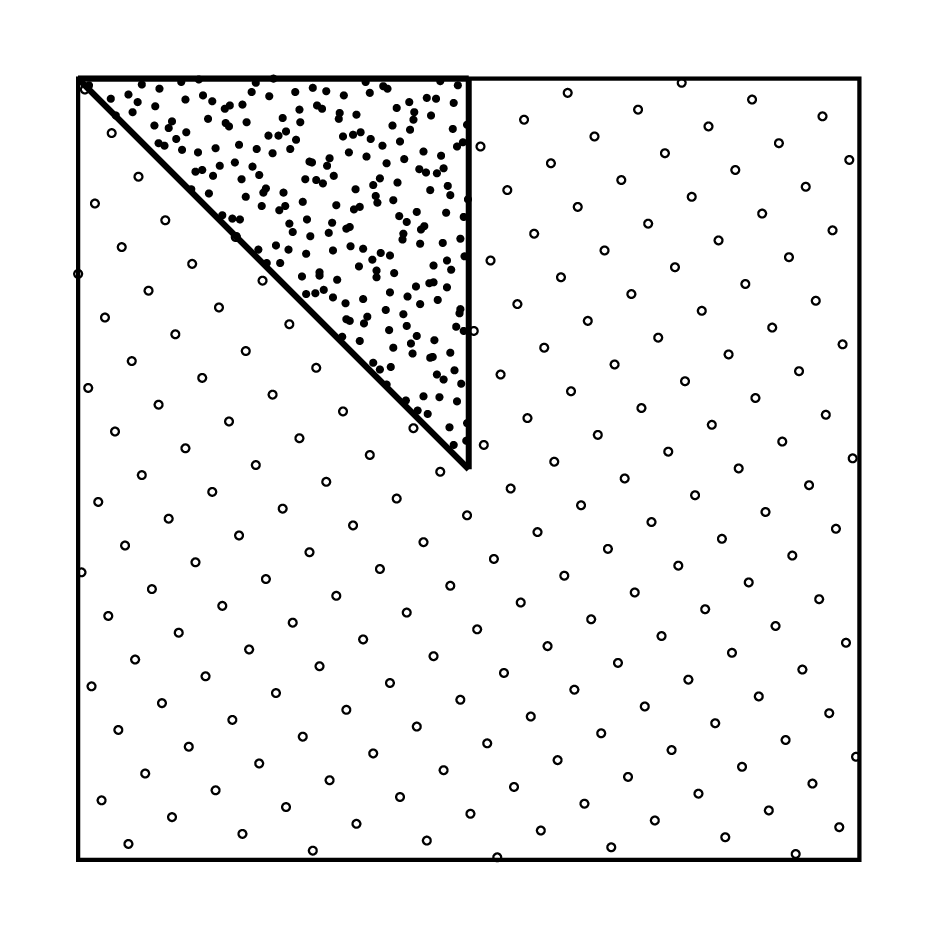}
    \end{minipage}
    \caption{Illustration of the importance of breaking the symmetry of the charge sampling grid. Both panels show a unit cell of a two-dimensional {\it P4m} lattice. The boldly outlined triangle denotes the fundamental domain. Empty circles indicate sampling points located outside the fundamental domain; filled circles show the corresponding positions folded back inside. Left panel: with a symmetric sampling grid, 7 out of 8 points are redundant. Right panel: breaking the grid symmetry yields an eightfold increase in sampling density within the fundamental domain.}
    \label{fig:symmetry}
\end{figure}

The role of symmetry in dual-space algorithms has long been debated (see \cite{palatinus2013charge} for a comprehensive review). Empirical evidence suggests symmetry constraints hinder convergence, leading to a common practice of relaxing them during phasing, only to re-impose them afterward. While the precise mechanism underlying this effect remains unclear, one plausible hypothesis is as follows. When the origin of the unit cell is placed at a Wyckoff position of highest site symmetry --- which is the typical convention --- the Cartesian sampling grid (\ref{eq:cartesian_grid}) is itself invariant under some or all of the space group operations. In the worst case, as illustrated in the left panel of Figure~\ref{fig:symmetry}, the grid may be fully invariant under the action of the entire space group. If symmetry constraints are enforced in this setting, charge density values at symmetry-equivalent grid points are forced to be identical, leading to substantial redundancy in the sampled data. By contrast, relaxing symmetry allows the algorithm to converge to a randomly shifted structure, thereby avoiding this redundancy.
\par
The toolkit adopts a different strategy to eliminate redundancy in charge density sampling by moving beyond conventional Cartesian grids. Although dual-space phasing methods require a periodic sampling of direct space to enable the use of the fast Fourier transform, this periodicity need not be restricted to Cartesian arrangements. In general, the sampling grid can be expressed as $\Lambda + \mathbf{t}$, where $\Lambda \supset L$ is a superlattice of the direct lattice $L$, and $\mathbf{t}$ is an arbitrary translation. Symmetry can then be explicitly broken by selecting $\Lambda$ to be invariant only under pure translations. The impact of this choice can be understood as follows. When symmetry is enforced, the charge density within the unit cell is fully determined by its values in a single fundamental domain of the space group. Phasing iterations can thus be interpreted as updates to the density within this domain, with sampled points folded back accordingly. As illustrated in the right panel of Figure~\ref{fig:symmetry}, using a fully asymmetric grid leads to a higher effective sampling density within the fundamental domain, reducing redundancy and potentially improving convergence.
\subsection{Efficient alias-free sampling}
\label{sec:antialias}
Aliasing \cite{amidror2013mastering_ch5} arises when the sampling grid is insufficiently dense, resulting in a loss of information in the sampled charge density. This effect is most naturally understood in reciprocal space. There, the dual of the sampling grid $\Lambda^*$ is a sublattice of the reciprocal lattice $L^*$. Let $M \subset L^*$ denote the set of wavevectors corresponding to measured reflections. Sampling is alias-free if
\begin{equation}
    k_1 - k_2 \in \Lambda^* \implies k_1 = k_2, \quad \forall\, k_1, k_2 \in M
    \label{eq:aliasing}
\end{equation}
In other words, translated copies of $M$ by $\Lambda^*$ must not overlap. Since the number of sampling points per unit cell equals the index $|L^* : \Lambda^*|$, minimizing this count corresponds to finding the densest periodic packing of copies of $M$.
\par
\begin{figure}[ht] %
    \begin{center}
        \includegraphics[width=0.6\textwidth]{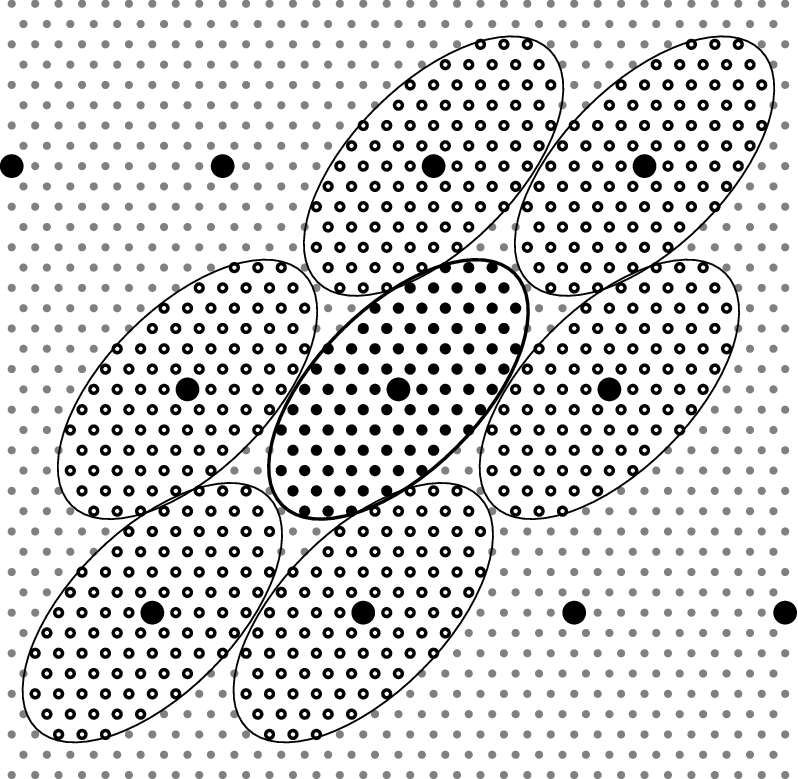} 
    \end{center}
    \caption{Illustration of the alias-free grid selection method. The figure shows reciprocal space. The lattice of small dots represents $L^*$, and the large black circles indicate its sublattice $\Lambda^*$. The small solid black dots denote the set $M$ of wavevectors for which amplitudes are known, while the thick black ellipse corresponds to the Löwner ellipsoid of $M$. The sublattice $\Lambda^*$ is constructed such that translated copies of $M$ (shown as hollow dots), shifted by vectors of $\Lambda^*$, do not overlap --- thus ensuring alias-free sampling of the charge density.} 
    \label{fig:superlattice}
\end{figure}
The algorithm used in the toolkit to construct near-optimal sampling grids approximates the set of measured reflections, $M$, by its Löwner ellipsoid \cite{henk2012lowner}. This approximation simplifies the problem of finding the densest periodic packing of translated copies of $M$ by reducing it to the classical problem of dense sphere packings \cite{conway2013sphere_ch1}. The algorithm proceeds as follows:
\begin{enumerate}
    \item Choose a basis $B$ for a lattice that densely packs unit-radius spheres in reciprocal space\footnote{For dimensions $\leq 8$, the toolkit uses the densest known periodic sphere packings from \cite{conway2013sphere_ch1}; for higher dimensions, it defaults to the $D_n$ lattice.}.
    \item Compute the Löwner ellipsoid\footnote{In the current implementation, the Löwner ellipsoid is approximated by a suitably scaled inertia ellipsoid of $M$.} of $M$ (i.e. the minimal-volume ellipsoid enclosing $M$), and let $Q$ be the quadratic form equal to 1 on this ellipsoid.
    \item \label{enum:random_matrix}Generate a random orthogonal matrix $U \in O(n)$, where $n$ is the dimension of $L$.
    \item \label{enum:rounding}Compute $A \in \mathrm{Mat}_n(\mathbb{Z})$ as the transformation matrix from the basis $B' = Q^{-1/2} U B$ to that of $L^*$, rounding entries to the nearest integer. This matrix defines the
    sampling grid by the formula
    \begin{equation}
        \Lambda = A^{-1} L
        \label{eq:Lambda}
    \end{equation}
    Due to the rounding effects, $\Lambda$ may not satisfy the condition (\ref{eq:aliasing}). If this occurs, repeat step 3, otherwise output $\Lambda$.
\end{enumerate}
An illustration of this construction in two dimensions is given in Figure \ref{fig:superlattice}. For aperiodic crystals and quasicrystals, the Löwner ellipsoid of $M$ provides a natural generalization of the {\em resolution sphere} \cite{ebashi1991handbook}. While the extent of $M$ in the physical dimensions of reciprocal space is inversely proportional to the radiation wavelength \cite{van2007incommensurate}, its extent in the perpendicular dimensions is limited only by the detector’s ability to register weak reflections. Imposing spherical symmetry on the region enclosing $M$ is therefore unjustified. In practice, when working with quasicrystals in programs such as SUPERFLIP, this issue is often mitigated by exploiting the inflation symmetry of the reciprocal lattice to make $M$ appear more ``round.'' However, this adjustment can only be made in discrete steps --- potentially large ones, such as for primitive icosahedral lattices, where the scale ratio between physical and perpendicular spaces can be adjusted only by powers of $\tau^6 \approx 17.9$. By contrast, the algorithm implemented in the present toolkit accommodates arbitrary anisotropy in the diffraction data.
\subsection{Cyclic sampling grids}
For non-Cartesian sampling grids, the transformation between diffraction amplitudes and charge density takes the form of a discrete Fourier transform between functions defined on the finite abelian groups $L^*/\Lambda^*$ and $\Lambda/L$. When these groups are cyclic, the transform can be performed by a one-dimensional fast Fourier transform (FFT). This special case is of particular interest from a computational standpoint for several reasons:
\begin{itemize}
    \item If $\Lambda$ is a cyclic superlattice of $L$ of index $N$, then it can be written as $\Lambda = L + \mathbb{Z} \cdot (\mathbf{v}/N)$ for some $\mathbf{v} \in L$. In this setting, the non-aliasing condition~(\ref{eq:aliasing}) corresponds to the requirement that the values of the scalar products $\mathbf{k} \cdot \mathbf{v}$ modulo $N$ for all $\mathbf{k} \in M$ should be distinct, which can be verified efficiently by sorting them.
    \item One-dimensional FFTs avoid strided memory access and are generally faster on modern hardware \cite{akin2014ffts}.
    \item Padding to the next power of two (or a product of small primes) in one dimension requires fewer additional sampling points than in the multidimensional case.
\end{itemize} 
\par
The abelian group $\Lambda/L$ is cyclic if and only if all but one of the invariant factors in the Smith normal form of the matrix $A$ in equation~(\ref{eq:Lambda}) are equal to 1. Since the algorithm described in Section~\ref{sec:antialias} involves a randomly generated orthogonal matrix at step~\ref{enum:random_matrix}, it is inherently non-deterministic. To ensure that $\Lambda/L$ is cyclic, an additional constraint is imposed at step~\ref{enum:rounding}, requiring the Smith normal form of $A$ to satisfy the above condition. In practice, this requirement is fulfilled in the majority of cases, consistent with the findings of \cite{wang2017smith}, which show that the probability of a random integer matrix having all but one invariant factor equal to 1 exceeds 0.84.
\section{Extensible architecture}
The toolkit is designed with a modular and extensible architecture, based on {\em dependency injection} via Julia's multiple dispatch mechanism. The core routine, \lstinline|do_phasing!|, serves as a generic driver for the phasing process. Its behavior is fully determined by the set of user-supplied interface objects passed as arguments. Three types of behavioral interfaces can be specified:
\begin{itemize}
    \item \textbf{Charge flipping algorithm} --- the core logic of the dual-space iteration (described in detail below).
    \item \textbf{Iteration control and user interaction} --- defines how the iteration process is managed. This interface supports pausing, resuming, and stopping the computation, and allows for interactive capabilities such as real-time visualization of the phasing progress. The toolkit includes two implementations: one for unattended (batch) execution, and one with an interactive graphical user interface.
    \item \textbf{Result output} --- defines how the final solution is saved. A default implementation is provided for exporting results in CSV format.
\end{itemize} 
\subsection{Charge flipping interface}
To implement a custom charge flipping algorithm, the user of the toolkit has to define a subtype of  \lstinline|AbstractPhasingAlgorithm| and supply new methods for the functions \lstinline|flip_charge!| and \lstinline|flip_amplitudes!|, which operate on instances of this subtype. Such instance may encapsulate arbitrary algorithmic context, such as threshold parameters, the number of iterations, or the complete history of the phase reconstruction.
\par
The two core functions of the interface are:
\begin{itemize}
    \item \lstinline|flip_charge!| --- applies a transformation to the charge density in direct space. It receives the array of charge density values and the interface object, and modifies the values in-place according to the internal state and parameters held by the object.
    \item \lstinline|flip_amplitudes!| --- updates the diffraction amplitudes. Due to symmetry constraints, each orbit of symmetry-related Bragg reflections is associated with a single amplitude value. The other effect of symmetry constraints is that the amplitudes are divided into two categories: those with arbitrary phases (complex-valued, denoted by subscript “c”) and those constrained to real values (denoted by subscript “r”).
\par    
    The function receives the following amplitude data:
    \begin{itemize}
        \item Normalized observed amplitudes: $|E^\text{obs}_\text{r}|$ and $|E^\text{obs}_\text{c}|$.
        \item Backprojected amplitudes: $E^\text{back}_\text{r}$ and $E^\text{back}_\text{c}$, computed from the charge density modified by the call to \lstinline|flip_charge!| and averaged over the action of the symmetry group. 
        \item Amplitude arrays to be updated: $E^\text{new}_\text{r}$ and $E^\text{new}_\text{c}$, to be set by the function.
    \end{itemize}
\end{itemize}
This interface is generic enough to allow for the implementation of phase retrieval algorithms beyond the traditional charge-flipping paradigm, such as the RRR algorithm \cite{elser2018benchmark}. Indeed, cyclically rearranging the main loop of Algorithm 1 in \cite{elser2018benchmark} by moving the "support size projection" step to the end allows for a clear separation between calls to \lstinline|flip_amplitudes!| and \lstinline|flip_charge!|. The auxiliary signal $\rho$ could then be naturally encapsulated within the corresponding \lstinline|AbstractPhasingAlgorithm| subtype.
\subsection{Reference implementation} \label{sec:reference_implementation}
The toolkit includes a reference implementation of the charge flipping interface, intended both as a usable default and as a template for custom algorithm development.
\par
In this implementation, the method \lstinline|flip_charge!| performs the following sequence of operations:
\begin{enumerate}
    \item \label{enum:rho0} Determine the threshold value $\rho_0$ such that a fixed fraction $\alpha$ of grid points have charge density values less than $\rho_0$, where $0 < \alpha < 1$ is a user-defined parameter.
    \item Apply the transformation
    \begin{equation}
        \rho \;\mapsto\;
        \begin{cases}
            \rho, & \text{if } \rho \ge \rho_0, \\
            2\rho_0 - \rho, & \text{if } \rho < \rho_0,
        \end{cases}
    \end{equation}
    effectively inverting the charge density values below the threshold about $\rho_0$.   
    \item Update the threshold fraction $\alpha$ by multiplying it by a fixed decrement factor (also user-specified).
\end{enumerate}
The initial value of $\alpha$ and the decrement factor are specified at the time of constructing the corresponding interface object.
\par
The accompanying implementation of \lstinline|flip_amplitudes!| is borrowed from the original charge flipping algorithm \cite{oszlanyi2004ab}. The update rules for the amplitudes are as follows:
\begin{eqnarray}
    E^\text{new}_\text{r} &=& \mathrm{sgn}(E^\text{back}_\text{r})|E^\text{obs}_\text{r}|\\
    E^\text{new}_\text{c} &=& \frac{E^\text{back}_\text{c}}{|E^\text{back}_\text{c}|}|E^\text{obs}_\text{c}|
\end{eqnarray}
\section{Benchmarks}
The toolkit was evaluated using two diffraction datasets for primitive icosahedral (PI) quasicrystals, with the symmetry constraint of the non-centrosymmetric PI symmetry group. The CdYb quasicrystal dataset (Takakura et al., 2007) comprised 5381 reflections, of which 11 were rejected as symmetry-equivalent to others, and a further 292 were excluded due to intensity uncertainties exceeding the measured values. Of the remaining 5078 independent reflections, 1380 lay in planes perpendicular to a twofold axis (real type, phase defined up to a sign), and 3698 were of the complex type (phase unrestricted). The entire set $M$ contained 522866 reflections counted with symmetry. The ZnMgTm quasicrystal dataset (Buganski et al., 2020) contained 3010 independent reflections, all retained, comprising 853 of the real type and 2157 of the complex type, with the total of 307368 reflections in the set $M$.
\subsection{Sampling grid density}\label{sec:density}
\begin{figure}[ht] %
    \begin{center}
        \includegraphics[width=0.8\textwidth]{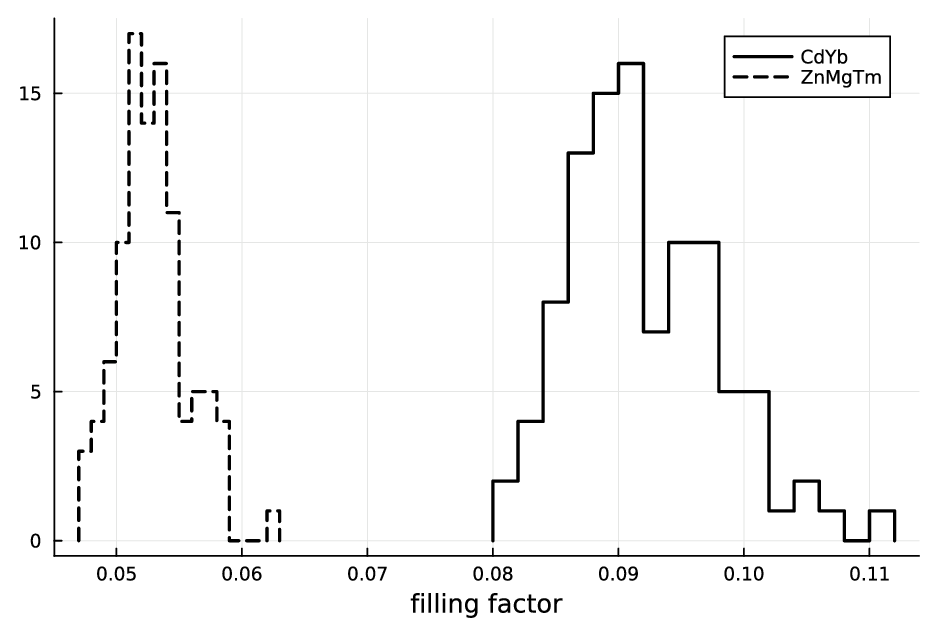} 
    \end{center}
    \caption{Ratio of the number of known amplitudes to the number of sampling grid points. Histograms were obtained from 100 independent runs of the algorithm described in Section~\ref{sec:antialias}, using the datasets of \protect\cite{takakura2007atomic} and \protect\cite{buganski2020atomic}.} 
    \label{fig:filling}
\end{figure}
For non-aliasing sampling, the number of grid points per unit cell must be at least equal to the number of measured reflections. In practice, a substantially larger number is required, since the image of the measured reflection set $M$ under the quotient map $L^* \to L^*/\Lambda^*$ only fills a small part of the quotient group. With the algorithm of Section \ref{sec:antialias}, the achievable filling factor is bounded above by the maximal density of periodic sphere packings in six dimensions,  $\pi^3/(48\sqrt{3}) \approx 0.373$ \cite{conway2013sphere_ch1}, yet the observed values are considerably lower (Fig. \ref{fig:filling}). Part of this shortfall can be attributed to the rounding at the step \ref{enum:rounding} in the algorithm, which is solely responsible for the run-to-run variation in the filling factor. From Fig. \ref{fig:filling}, this rounding accounts for approximately 20–30\% of the total deficit. The remaining deficit is attributable to missing data i.e. the points of $L^*$ lying within the Löwner ellipsoid of $M$ for which diffraction intensities were not measured. Notably, the dataset of \cite{buganski2020atomic} exhibits a markedly higher proportion of missing data compared to that of \cite{takakura2007atomic}. 
\subsection{Performance metrics}
\begin{table}[ht]
    \caption{Execution time per iteration and peak memory usage (maximum resident set size, MaxRSS) for the Julia process. Benchmarks were performed using diffraction data from the icosahedral quasicrystals CdYb \protect\cite{takakura2007atomic} and ZnMgTm \protect\cite{buganski2020atomic}, in two execution modes: unattended batch processing and interactive execution with a graphical user interface (GUI). In all cases, the reference implementation of the charge flipping algorithm was used. Measurements were obtained on a system with an AMD Ryzen 7 3700X CPU (8 cores, 2.2 GHz base clock), 64 GB DDR4-3600 RAM (2-channel). Software configuration: Julia v1.11.6 with FFTW.jl v1.9.0, Linux 5.10.0-29-amd64.} 
    \smallskip
    \begin{center}
        \begin{tabular}{lccr}
            Task    & Time per iteration (s)              & MaxRSS (MiB)     \\  
            \hline          
            CdYb unattened      & 0.393            & 1854      \\
            CdYb GUI      & 0.609            & 2544      \\
            ZnMgTm unattended      & 0.184            & 1677      \\
            ZnMgTm GUI      & 0.351            & 2157      \\
        \end{tabular}
    \end{center}
    \label{tab:performance}
\end{table}
The performance benchmarks are summarised in Table \ref{tab:performance}. Profiling indicates that the largest share of execution time is spent in step \ref{enum:rho0} of the algorithm described in Section \ref{sec:reference_implementation}. This is hardly surprising, as the default sorting procedure employed at this step is single-threaded, in contrast to the highly optimised multithreaded FFT routines. Note also the substantial overhead introduced by the graphical user interface, which can slow execution by up to a factor of two.
\subsection{Convergence behaviour}
\begin{figure}[ht] %
    \begin{center}
        \includegraphics[width=0.8\textwidth]{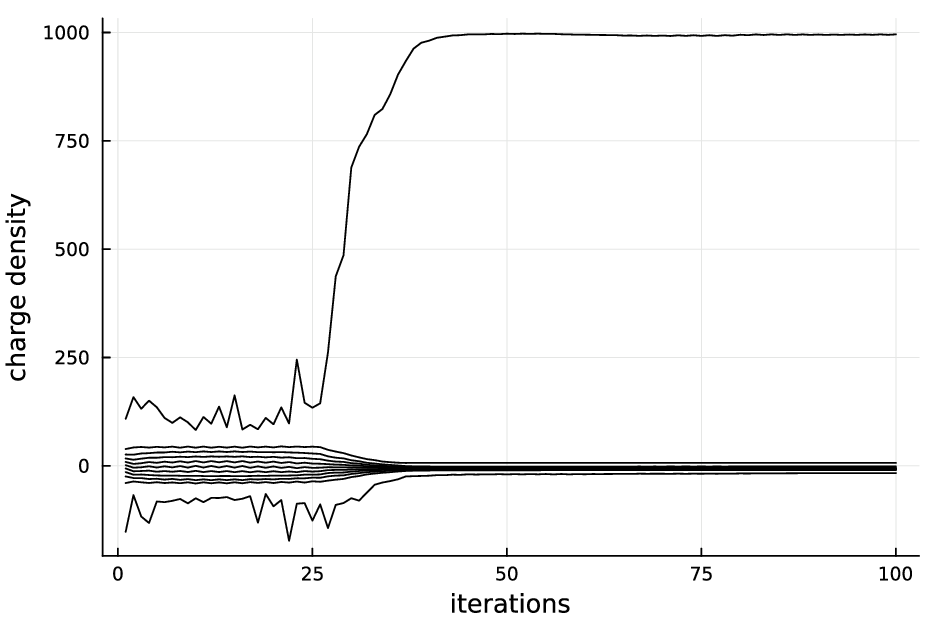} 
    \end{center}
    \caption{Evolution of the minimum, maximum, and all deciles of the charge density values on the sampling grid during the execution of the reference implementation of the algorithm, using the data of \protect\cite{takakura2007atomic}. The run was initialized with $\alpha = 0.8$ and a decrement factor of 0.995.} 
    \label{fig:deciles}
\end{figure}
\begin{figure}[ht] %
    \begin{center}
        \includegraphics[width=0.8\textwidth]{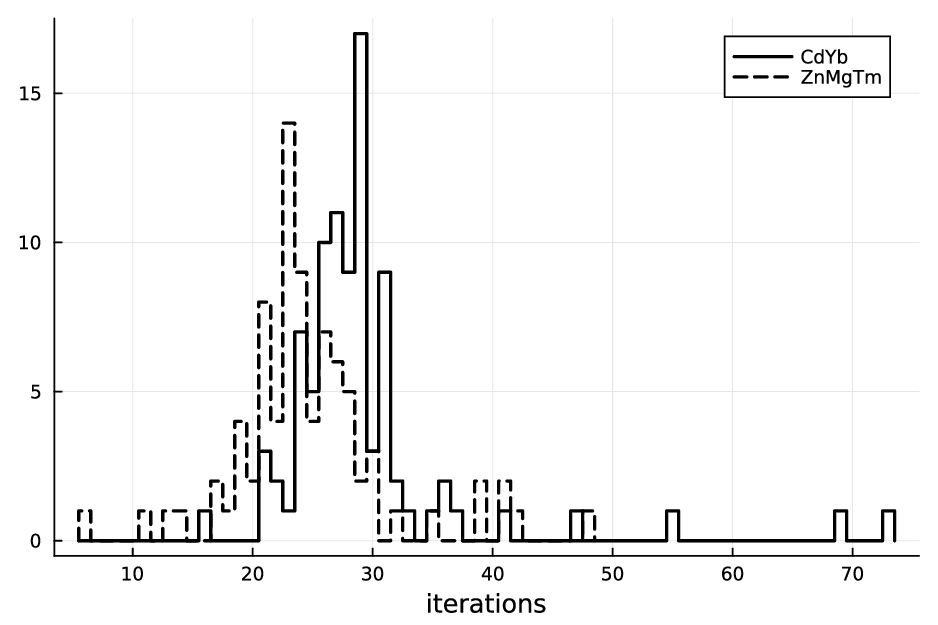} 
    \end{center}
    \caption{Number of iterations required for convergence of the reference implementation of the algorithm (decrement factor = 0.99). Convergence is defined as the point at which the ratio between the deviations of the maximum and minimum charge density from their median value exceeds 50. Each curve is based on 100 independent runs, with a maximum of 100 iterations per run. The solid line corresponds to the CdYb quasicrystal data of \protect\cite{takakura2007atomic}, for which convergence was not achieved in 10 cases. The dashed line corresponds to the ZnMgTm quasicrystal data of \protect\cite{buganski2020atomic}, for which convergence was not achieved in 17 cases.} 
    \label{fig:convergence}
\end{figure}
\begin{figure}[ht] %
    \begin{center}
        \includegraphics[width=0.8\textwidth]{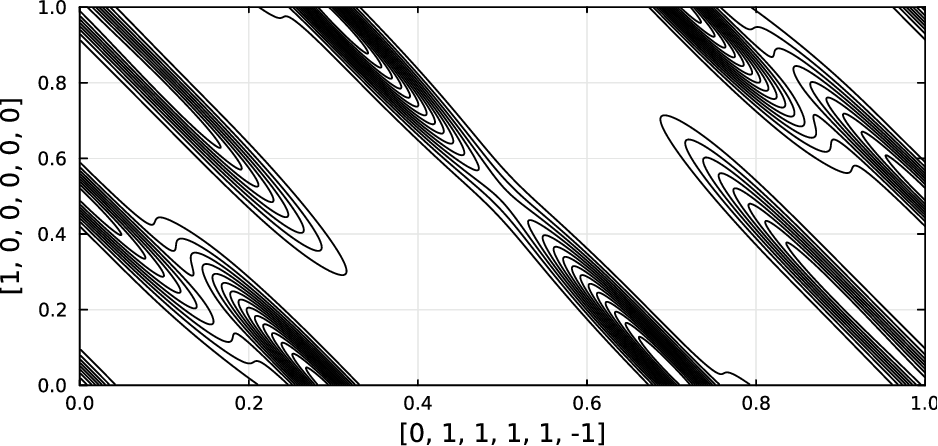} 
    \end{center}
    \caption{Contour plot of a section of the charge density of the CdYb icosahedral quasicrystal in the plane invariant under fivefold rotations. The axis labels correspond to the direct lattice vectors spanning this plane. Data from \protect\cite{takakura2007atomic}.} 
    \label{fig:cdyb_fivefold}
\end{figure}
Since charge-flip algorithms are designed with the implicit objective of maximizing the global minimum of the charge density, they tend to converge to density distributions with large flat valleys and narrow, sharp peaks. This makes the quantiles of the charge density a natural quantity to analyze to assess the algorithm’s convergence. Figure~\ref{fig:deciles} shows the evolution of the deciles of the charge density together with its extremal values during iterations. A drastic decrease in the distance between the minimal value and all deciles, as well as a sharp increase in the maximal value, can be observed at the moment when the solution is found.
\par
Another useful metric for evaluating charge-flipping algorithms is the success rate (the proportion of cases in which the algorithm converges to a meaningful solution), as well as the speed of convergence. In this benchmark, the convergence point was defined as the moment at which the ratio of the deviation of the extrema from the median charge density exceeded 50. Figure~\ref{fig:convergence} presents the results for 100 independent runs for both test cases. The success rates after 100 iterations for the CdYb and ZnMgTm datasets were 90\% and 83\%, respectively, with convergence occurring in most cases between the 20th and 30th iterations.
\subsection{Visualization of results}
To visualize the solution, the toolkit provides a means of generating two-dimensional maps of the charge density restricted to arbitrary rational planes. Figure~\ref{fig:cdyb_fivefold} shows such a map for the CdYb quasicrystal, for the plane spanned by the vectors $[1,0,0,0,0,0]$ and $[0,1,1,1,1,-1]$. This plane contains the fivefold axes in both the physical and perpendicular spaces, and is therefore invariant with respect to fivefold rotations. 
\section{Discussion and conclusions}
Running the toolkit on real experimental data produced some unexpected results. Notably, the rudimentary implementation of the \lstinline|flip_charge!| method (Section~\ref{sec:reference_implementation}) proved to be surprisingly efficient. A plausible explanation is that the gradual reduction of the charge-flipping threshold $\rho_0$ allows the algorithm time to ``find the entrance'' to the convergence valley. It is well established that correct setting of this threshold is critical for successful convergence \cite{palatinus2007superflip, palatinus2013charge}. With a slow decrease of $\rho_0$, the algorithm will inevitably perform several iterations with the threshold close to its unknown optimal value, sufficient to initiate convergence. Furthermore, continued reduction of $\rho_0$ at later stages suppresses small “ripples” at the bottoms of valleys in the charge-density distribution, thus removing the need for any additional refinement step.
\par
Another unusual observation was the markedly low filling factor of the reciprocal grid $L^*/\Lambda^*$ and the pronounced variations in this factor, attributable to the rounding step in the sampling grid generation algorithm (Section~\ref{sec:antialias}). This serves as a reminder that the geometry of high-dimensional spaces is often counterintuitive. In particular, estimating the extent of missing data is difficult without actually running the program.
\par
We now turn to possible improvements to the toolkit. The most significant would address the treatment of missing and unmeasured data. For periodic crystals, this corresponds to diffraction data missing for reciprocal-lattice points lying within the {\em resolution sphere} \cite{palatinus2013charge}. A natural generalization of this concept for aperiodic crystals and quasicrystals is the Löwner ellipsoid of the set $M$ (or, in cases involving extrapolating low-resolution data, any other ellipsoid containing $M$ and possessing the symmetry of the structure). In its current implementation, the toolkit does not supply amplitudes for the missing data to the \lstinline|flip_amplitudes!| function, effectively setting them to zero. Modifying this behaviour would entail a breaking change to the toolkit’s API. Nevertheless, such a modification appears essential given the importance of properly addressing the missing-data problem.
\par
Two additional modifications could potentially enhance the computational performance of the program. The first aims to mitigate the reduction of the filling factor of the reciprocal grid $L^*/\Lambda^*$ caused by the rounding operation. The effect of rounding depends on the chosen basis of $L^*$. In the current implementation, rounding is performed in the same basis in which the diffraction data are defined. To reduce rounding errors, it would be preferable to use a basis reduced with respect to the quadratic form on $L^*$ that takes the value 1 on the Löwner ellipsoid of $M$. Such a basis reduction could be achieved using the LLL algorithm \cite{lenstra1982factoring}. The second proposed improvement concerns the time-consuming computation of the $\alpha\text{-th}$ quantile $\rho_0$ of the charge density in step \ref{enum:rho0} of the algorithm described in Section~\ref{sec:reference_implementation}. As only an approximate value of $\rho_0$ is required, it could be estimated from the density values at a small, randomly selected fraction of the grid points.
\par
The final potential improvement concerns the user experience with the toolkit. In the current implementation, the user is responsible for defining the regularizing form factors (see Section~\ref{sec:formfactors}). However, these factors should depend only on the sampling grid $\Lambda$ and on the set of known reflections $M$. They could therefore be optimized for an artificial structure containing a single $\delta\text{-peak}$ at an arbitrary position within the unit cell. Such optimization constitutes a well-posed mathematical problem, and solving it would relieve the user of this task.
\par
In conclusion, the present work demonstrates that crystallographic symmetry can be applied to the charge density during charge-flipping iterations without compromising computational performance. Benchmarks indicate that the toolkit is capable of solving the phase problem with datasets containing as many as half a million reflections within only tens of seconds on high-end desktop systems. Such performance facilitates rapid development of new variants of charge-flipping algorithms.
\section*{Acknowledgments}
This work benefited greatly from many discussions with my dear friend and colleague, the late André Katz, whose wisdom and friendship are deeply missed.

\bibliographystyle{unsrt}
\bibliography{open_phaser_arxiv}
\end{document}